# Running coupling constant and correlation length from Wilson loops

Massimo Campostrini, Paolo Rossi, and Ettore Vicari,
Dipartimento di Fisica dell'Università and I.N.F.N., I-56126 Pisa, Italy

We consider a definition of the QCD running coupling constant $\alpha(\mu)$ related to Wilson loops of size $r \times t$ with arbitrary fixed $t/r$. The schemes defined by these couplings are very close to the $\overline{\rm MS}$ scheme for all values of $t/r$; in the $t/r \to \infty$ limit, the "$q\bar{q}$ force" scheme is recovered. We propose a definition of correlation length, also related to Wilson loops, which can be applied to the Monte Carlo evaluation of $\alpha(\mu)$ up to very large momentum scales by use of finite-size scaling techniques.

## 1. Introduction

A precise determination of $\Lambda_{\overline{\rm MS}}$ in units of a physical mass scale is an important goal of lattice QCD; it involves an accurate determination of $\alpha_{\overline{\rm MS}}(\mu)$ at some large momentum scale $\mu$, where perturbation theory is reliable and accurate. On phenomenological grounds, a direct computation of $\alpha_{\overline{\rm MS}}(\mu \sim m_{Z_0})$ would be most welcome.

In the last few years, considerable progress has been achieved towards this goal [1–6]; however, the tasks of performing lattice calculations at large momentum scales turns out to be quite hard and therefore some kind of perturbative extrapolation has to be employed.

In order to reach very large momentum scales without perturbative extrapolations, the computation of $\alpha(\mu)$ requires finite-size scaling techniques, which permit to reach very small distances without the need of very large lattices. The power of this approach has been illustrated recently by a study of two-dimensional spin models in the extremely small distance region [7].

This kind of technique requires a careful choice both of the observable and of the correlation length $\xi$: they must be easy to measure to high precision on the lattice, and must enjoy well-defined finite-size properties. Furthermore in our case we require $\alpha(\mu)$ to be close to $\alpha_{\overline{\rm MS}}$, to reduce errors due to neglected orders in the perturbative conversion to the $\overline{\rm MS}$ scheme.

In the following we will present new definitions of $\alpha(\mu)$ and $\xi$ addressing the above points.

## 2. Coupling constant

A useful definition of running coupling constant is derived from the static quark-antiquark force $F(r)$ by the relationship

$$F(r) = -c_F \frac{\alpha_{q\bar{q}}(1/r)}{r^2}, \qquad (1)$$

where $c_F = (N^2-1)/(2N) = \frac{4}{3}$. (We prefer to write $\alpha_{q\bar{q}}$ with a momentum scale dependence, rather then with a length scale dependence, as often seen in the literature).

This definition is well suited for as lattice computation [4,5], but is unsuitable for our purposes, since the computation of $F(r)$ requires a $t \to \infty$ limit procedure in order to evaluate $F(r)$, and it is not clear how to implement it on a finite lattice. We therefore propose to modify the definition (1) to avoid the limiting procedure.

The weak coupling expansion of the Creutz ratio is

$$\chi(r,t) \equiv \frac{\partial^2 \ln W(r,t)}{\partial r \partial t}$$
$$= -c_F A(x) \frac{\alpha}{r^2} (1+O(\alpha)), \qquad (2)$$

where $x = t/r$ (we can choose $x \geq 1$ without loss of generality), and

$$A(x) = \frac{2}{\pi}\left[\arctan(x) + \frac{1}{x} + \frac{\arctan(1/x)}{x^2}\right]. \qquad (3)$$

The renormalization properties of the Wilson loop operator [8,9] allow the definition of a running coupling constant $\alpha_x(1/r)$ parametrized by $x$:

$$\chi(r,xr) = -c_F A(x) \frac{\alpha_x(1/r)}{r^2}. \qquad (4)$$

Table 1
The quantity $R = \ln(\Lambda_x/\Lambda_{\overline{MS}})$ as a function of $x$ and of the number of flavours $N_f$.

| $x$ | $N_f = 0$ | $N_f = 2$ | $N_f = 4$ |
|---|---|---|---|
| 1 | −0.20758 | −0.27875 | −0.37270 |
| 5/4 | −0.11177 | −0.18145 | −0.27343 |
| 4/3 | −0.08975 | −0.15852 | −0.24930 |
| 3/2 | −0.05567 | −0.12242 | −0.21053 |
| 5/3 | −0.03131 | −0.09603 | −0.18146 |
| 7/4 | −0.02178 | −0.08554 | −0.16970 |
| 2 | −0.00069 | −0.06189 | −0.14266 |
| 3 | 0.03187 | −0.02341 | −0.09637 |
| 4 | 0.03979 | −0.01317 | −0.08308 |
| 10 | 0.04557 | −0.00500 | −0.07175 |
| ∞ | 0.04691 | −0.00324 | −0.06945 |

Since by definition

$$\lim_{t \to \infty} \chi(r, t) = F(r), \qquad (5)$$

we have

$$\lim_{x \to \infty} \alpha_x(1/r) = \alpha_{q\bar{q}}(1/r). \qquad (6)$$

We computed the quantity $R(x) = \ln(\Lambda_x/\Lambda_{\overline{MS}})$, where $\Lambda_x$ is the $x$-dependent $\Lambda$-parameter associated with $\alpha_x(\mu)$, by standard perturbation theory [10]. Here we only present a summary of the results in Table 1, for $N = 3$ and different choices of $x$ and $N_f$. It is interesting to notice that $R(x)$ is reasonably small for all $x \geq 1$; in the case $x = 2, N_f = 0$, $R$ is negligible, i.e., $\Lambda_2 \cong \Lambda_{\overline{MS}}$.

## 3. Finite size scaling

On the lattice the quantity

$$\phi(r, x) = \chi_C\left(\frac{r}{a}, \frac{xr}{a}\right), \qquad (7)$$

where $\chi_C$ is the usual Creutz ratio, is a natural estimator of $\chi(r, xr)$. By measuring $\phi(r, x)$ at different scales $r$ keeping $x$ fixed, one may easily extract the corresponding running coupling $\alpha_x(1/r)$.

In the scaling region the following finite size scaling relations hold:

$$\xi_L(\beta) \simeq f_\xi\left(\frac{L}{\xi_L}\right)\xi_\infty(\beta), \qquad (8)$$

and

$$\phi(\beta, r, x, L) \simeq f_\phi\left(x, \frac{L}{r}, \frac{L}{\xi_L}\right)\phi(\beta, r, x, \infty). \qquad (9)$$

Finite size scaling functions like $f_\xi$ and $f_\phi$ can be reconstructed by performing simulations on relatively small lattices, and we should be able to keep $a \ll r$ even when $r$ is very small (in physical units).

## 4. Correlation length

A crucial ingredient in finite size methods is a suitable definition of correlation length $\xi$, which can be measured to high precision in a Monte Carlo simulation. In two-dimensional spin models, very good results were obtained using $\xi$ defined from the second moment of a correlation function [7,11]. We propose a similar definition of $\xi$ for (confining) gauge theories, stemming from a correlation function which is easily constructed from Wilson loop and has the correct properties:

$$Y(r, t) = \frac{W(r, t)}{W\left(\frac{1}{2}r, \frac{1}{2}t\right)^2}. \qquad (10)$$

The perimeter-term divergence cancels in the ratio, leaving a distance-independent multiplicative renormalization, and the area law insures exponential fall down at large distances.

From $Y(r, t)$ we can define a second moment type correlation length:

$$\xi_Y^2 = \frac{1}{2}\frac{\int_0^\infty dr \int_r^{\varkappa r} dt\, Y(r, t)\, rt}{\int_0^\infty dr \int_r^{\varkappa r} dt\, Y(r, t)}, \qquad (11)$$

where $\varkappa$ is a free parameter ($\varkappa > 1$), which can be chosen to optimize the measurement (we introduced $\varkappa$ to avoid potential problems with "very thin" loops which are present in the integral for $\varkappa = \infty$).

In the case of an exact area law for the Wilson loop, we would get $\xi_Y^2 = 1/\sigma$.

The measurement of $\alpha_x$ and $\xi_Y$ up to a large scale $\mu$ leads to a direct determination of the $\beta$-function of the SU(3) lattice gauge theory and of the adimensional quantity $\xi_Y \Lambda_{\overline{MS}}$.

This quantity still needs to be converted to a more phenomenological scale, such as $r_0$ [12]; but this is a rather minor problem, since it involves only a measurement of $\xi_Y/r_0$, which can be performed at the values of $\beta$ of our choice.

## 5. Summary and outlook

We presented a definition of $\alpha(\mu)$ in terms of Wilson loops which is very close to $\alpha_{\overline{\text{MS}}}(\mu)$. This definition can be applied without changes in the case of QCD with dynamical fermions.

Moreover, we presented a definition of correlation length $\xi_Y$ in pure gauge QCD which is similar to "second moment" correlation lengths in spin models. (In the case of unquenched QCD a "physical" definition such as the inverse nucleon mass can be used.)

Both quantities do not require any extrapolation to long distances and enjoy well-defined finite-size scaling properties. Therefore they are well suited for a Monte Carlo study reaching a very large momentum scale by use of finite size techniques.